\documentclass[a4paper,11pt]{article}
\usepackage{pos}
\usepackage{float}
\usepackage{caption}
\usepackage{subcaption}

\title{Charmonium spectroscopy with optimal distillation
profiles}

\author*[a]{Juan Andrés Urrea-Niño}
\author[a]{Jacob Finkenrath}
\author[a]{Roman Höllwieser}
\author[a]{Francesco Knechtli}
\author[a]{Tomasz Korzec}
\author[b]{Michael Peardon}

\affiliation[a]{Dept. of Physics, Bergische Universität Wuppertal,\\
  Gaußstraße 20, 42119 Wuppertal, Germany}

\affiliation[b]{School of Mathematics, Trinity College Dublin,\\
Dublin 2, Ireland}

\emailAdd{urreanino@uni-wuppertal.de}
\emailAdd{finkenrath@uni-wuppertal.de}
\emailAdd{hoellwieser@uni-wuppertal.de}
\emailAdd{knechtli@uni-wuppertal.de}
\emailAdd{korzec@uni-wuppertal.de}
\emailAdd{mjp@maths.tcd.ie}

\abstract{We use the method of optimal distillation profiles to compute the low-lying charmonium spectrum in an $N_f = 3+1$ ensemble at the $SU(3)$ light flavor symmetric point ($m_{\pi} \approx 420$ MeV), physical charm quark mass and lattice spacing $a\approx 0.0429$ fm. The spectrum and mass splittings display good agreement with their values in nature and the statistical errors are comparable, if not smaller, than those of state-of-the-art lattice calculations. We also present first results on the mixing of charmonium with glueballs and light hadrons obtained in a similar $N_f = 3+1$ ensemble but at larger pion mass.}

\FullConference{The 40th International Symposium on Lattice Field Theory (Lattice 2023)\\
July 31st - August 4th, 2023\\
Fermi National Accelerator Laboratory\\}

\begin{document}
\maketitle

\section{Introduction}
\noindent The method of optimal distillation profiles introduced in \cite{Urrea} has been used to define meson operators with large overlaps onto energy eigenstates of interest, significantly improving on standard distillation \cite{Peardon}, and to study glueball-charmonium mixing \cite{Urrea-Proc2022} and static potentials \cite{Roman-Static}. The goal of this work is to use it to study the charmonium spectrum and its mixing with glueballs and light mesons in $N_f = 3 + 1$ gauge ensembles. The significant suppression of excited-state contamination at small time separations yielded by the optimal profiles is expected to improve the signal of quark-disconnected contributions, which are necessary to study iso-scalar meson operators yet heavily affected by a signal-to-noise problem. Two different pion masses $\left( \frac{m_{\eta_c}}{m_{\pi}} \approx 3, 7\right)$ are used, which change the decay channels for glueballs and two-particle states. A first step to study these decay dynamics is also done in this work by mapping out the energy spectrum based on single-particle operators.

\section{Methods}
\noindent Two ensembles of $N_f = 3 + 1$ clover improved Wilson fermions, Lüscher-Weisz gauge action, open boundary conditions in time at the $SU(3)$ light flavor symmetric point are used in this work \cite{Roman-Ensembles, Fritzsch2018}. The first one, denoted as B, has size $48^3 \times 144$, $\beta = 3.43$, lattice spacing $a\approx 0.043$ fm and $m_{\pi} \approx 420$ MeV. The second one, denoted as A1-heavy, has size $32^3 \times 96$, $\beta = 3.24$, $\frac{m_{\eta_c}}{m_{\pi}} = 3$ and lattice spacing $a \approx 0.068 $ fm determined from the spin-singlet splitting $m_{h_c} - m_{\eta_c}$. The latter ensemble is particularly useful for meson-glueball mixing since the two-pion decay threshold is significantly raised while the former ensemble is particularly useful for mapping the charmonium spectrum thanks to the light quark masses being tuned to their sum in nature \cite{Roman-Proc}. The observable of interest for these ensembles is the temporal correlation matrix
\begin{align}
C(t) &= \begin{pmatrix}
C_{cc}(t)  & C_{cl}(t) & C_{cg}(t)\\
C_{lc}(t) & C_{ll}(t)  & C_{lg}(t)\\
C_{gc}(t) & C_{gl}(t) & C_{gg}(t)
\end{pmatrix}.
\label{eqn:CorrMatrix}
\end{align}
whose entries involve different types of operators. The upper left $2\times 2$ block contains only iso-singlet mesonic operators: $C_{q_1 q_2}(t)$ is the correlation $ \left \langle \bar{q}_1(t) \Gamma q_1(t) \cdot \bar{q}_2(0) \tilde{\Gamma} q_2(0) \right \rangle$, where $\tilde{\Gamma} = \gamma_0 \Gamma^{\dagger} \gamma_0$ and $q_1, q_2 \in \{ c,l \}$. The off-diagonal terms $C_{cl}(t)$, $C_{lc}(t)$ of this block contain information about flavor-mixing between charmonium and light mesons, with the case of mixing between $\eta_c $ and $\eta^{\prime}$ being of particular interest for this work. The remaining elements $C_{qg}(t)$, $C_{gq}(t)$ outside of the $2\times 2$ block contain information about mixing of the charmonium and light iso-singlets with gluonic operators, which in this work are built from the eigenvalues of the 3D gauge covariant lattice Laplacian operator \cite{Morningstar2013}. Each of the $9$ entries of the correlation matrix in Eq~\eqref{eqn:CorrMatrix} are by themselves a matrix since one can use multiple mesonic and gluonic operators with the same quantum numbers. For purely mesonic correlations, i.e $C_{q_1q_2}(t)$, the entries of these matrices are given by
\begin{align}
C_{q_1 q_2}(t)_{mn} &= \delta_{q_1 q_2}\left \langle - \text{Tr}\left( \Phi_m[t] \tau_{q_1}[t,0] \bar{\Phi}_n[0] \tau_{q_2}[0,t]  \right) \right \rangle_{\text{gauge}}  \\
&+ \sqrt{N_{q_1}} \sqrt{N_{q_2}} \left \langle  \text{Tr}\left( \Phi_m[t] \tau_{q_1}[t,t]  \right) \text{Tr}\left( \bar{\Phi}_n[0] \tau_{q_2}[0,0] \right)    \right \rangle_{\text{gauge}} \nonumber.
\end{align}
where $N_q$ denotes the degeneracy of the flavors ($N_c = 1$, $N_l = 3$) and the indices $m$, $n$ go from $0$ to $N_B -1$, the number of different meson operators chosen in this work to be the same for charm and light flavors. The vacuum expectation value contributions involving $\left \langle  \text{Tr}\left( \Phi_m[t] \tau_{q_1}[t,t]  \right)\right \rangle_{\text{gauge}}$ are explicitly subtracted only in the $0^{++}$ symmetry channel. The modulated elementals $\Phi_m[t]$ for a fixed choice of $\Gamma$ have entries
\begin{align}
\Phi_m[t]_{\substack{ij\\ \alpha \beta}} &= f_m\left( \lambda_i[t], \lambda_j[t]  \right) v_i[t]^{\dagger} \Gamma_{\alpha \beta} v_j[t]
\end{align}
for a given choice of meson profiles $f_m\left(  \lambda_i[t], \lambda_j[t] \right)$, $m = 0,...,N_B-1$. $\bar{\Phi}_m[t]$ is defined in the same manner but using $\tilde{\Gamma} = \gamma_0 \Gamma^{\dagger} \gamma_0$ and $\tau_q[t_1,t_2] = V[t_1]^{\dagger} D_q^{-1} V[t_2]$ is the perambulator for the quark flavor $q$. The matrix $V[t]$ has $4\times N_v$ columns corresponding to the $N_v$ Laplacian eigenvectors placed into each of the 4 possible spin indices, making it block diagonal in spin. The value of $N_v$ used for the charm and light perambulators can be different, which can be preferable due to the inversions being more expensive for the light quarks. For purely gluonic correlations and those involving gluonic-mesonic mixing, the entries of the matrices are given by
\begin{align}
C_{qg}(t)_{mb} &= \sqrt{N_q} \left \langle \text{Tr}\left( \Phi_m[t] \tau_{q}[t,t] \right) G^{R}_b(0) \right \rangle_{\text{gauge}}\\
C_{gg}(t)_{ab} &= \left \langle G_a^{R}(t) G_b^{R}(0)  \right \rangle_{\text{gauge}},
\end{align}
where $G^{R}_a(t)$, $a = 0,...,N_G - 1$, are a set of $N_G$ glueball operators chosen to transform according to the same irrep $R$ as the meson $\bar{q}\Gamma q$. For the scalar channel the glueball operators are built from the sum of the $N_v$ Laplacian eigenvalues at a given time weighted by different profiles, similar to the mesonic elementals. Other glueball operators built from spatial Wilson loops as described in \cite{PeardonGlueballs, Berg} including several loop shapes and levels of APE smearing \cite{APE} were tried but the ones from the eigenvalues yielded the best signal. Different number of mesonic and gluonic operators are used in this work and therefore $C_{gg}(t)$ does not have the same size as $C_{qq}(t)$ and $C_{gq}(t)$ is a rectangular matrix. \\
\vspace*{0.1in}

\noindent The energies of the different states of interest are extracted by solving a generalized eigenvalue problem (GEVP) \cite{Luscher, Blossier} given by
\begin{align}
\tilde{C}(t) u_n(t,t_0) &= \rho_{n}(t,t_0) \tilde{C}(t_0) u_n(t,t_0),
\end{align} 
where the matrix $\tilde{C}(t)$ is obtained by projecting the correlation matrix $C(t)$ onto the singular vectors with largest singular values of $C(t_0)$. This keeps the contributions of orthogonal operators with good overlap onto the low-lying states and makes the problem better conditioned against statistical noise \cite{Balog, Niedermayer}. The effective masses of the $n$-th state are then extracted as
\begin{align}
am_{\text{eff}}^{n} = \ln\left( \frac{\rho_n(t,t_0)}{\rho_n(t+a,t_0)} \right).
\end{align}
To systematically study the effects of the different entries of the correlation matrix in Eq.~\eqref{eqn:CorrMatrix}, one can solve the GEVP starting not with the full correlation matrix but only with sub-blocks of it. For example, taking only $C_{cc}(t)$ allows to study the charmonium spectrum but neglects possible mixing with light hadrons or gluonic operators. Taking the upper left $2\times 2$ sub-block in Eq.~\eqref{eqn:CorrMatrix} allows to study the charmonium and light spectrum including their mixing, yet it neglects mixing with gluonic operators. Different combinations are treated in this work. When using only mesonic operators, the corresponding optimal profile for the $n$-th state is given by 
\begin{align}
\tilde{f}^{(n)}\left( \lambda_i[t], \lambda_j[t] \right) &= \sum_{k} u_n^{(k)}(t_1,t_0) f_{k}\left( \lambda_i[t], \lambda_j[t] \right),
\end{align}
where $u_n^{(k)}(t_1,t_0)$ denotes the $k$-th entry of the vector $u_n(t_1,t_0)$ and $t_1$ is a value of time chosen such that excited-state contamination is sufficiently suppressed.

\section{Spectrum results}
\begin{figure}[H]
\centering
\includegraphics[width=0.6\textwidth]{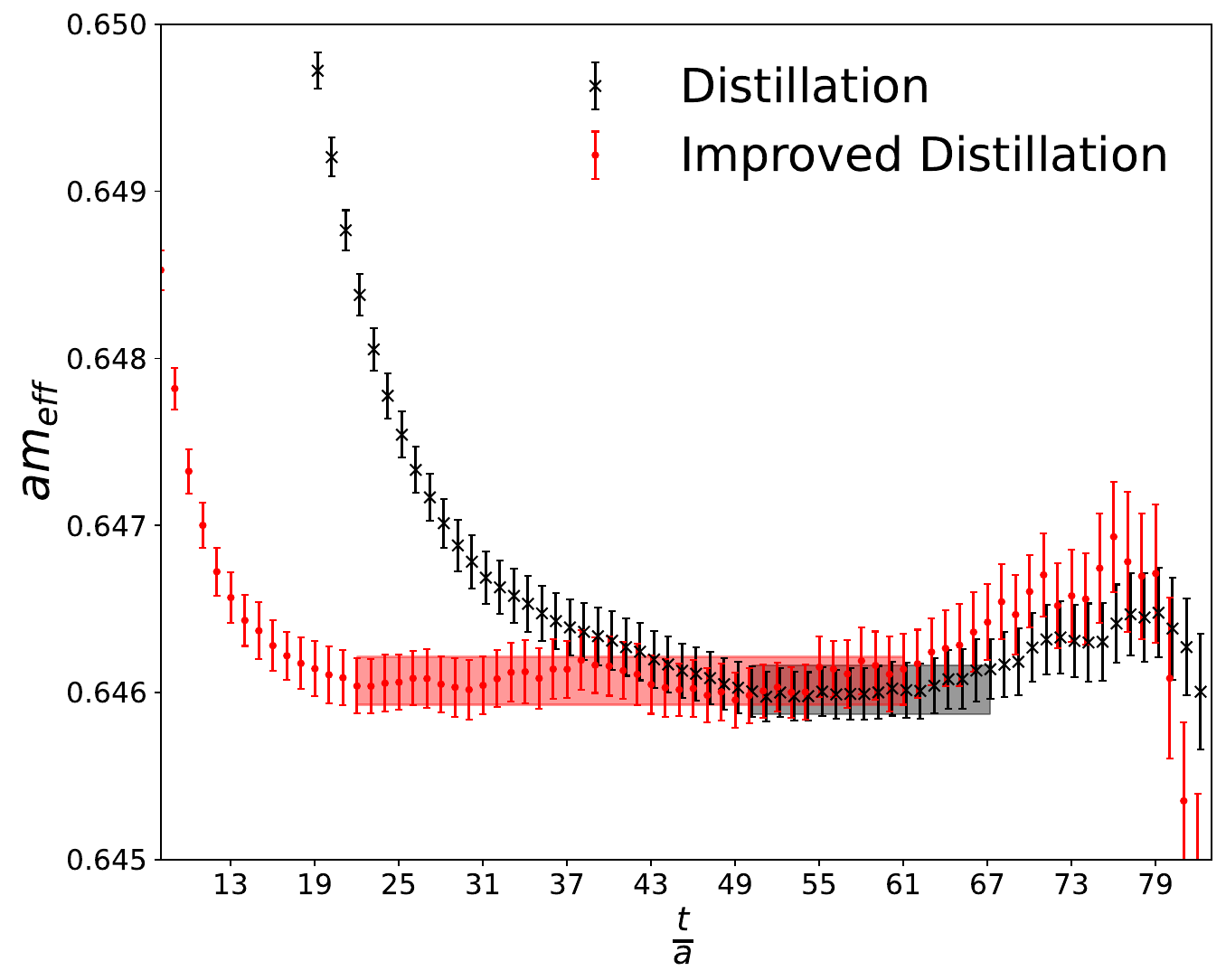}
\caption{Ground state effective masses of the $\Gamma = \gamma_5$ charmonium operator using only the quark-connected contribution to the correlation function using standard distillation and the improved variant with optimal meson distillation profiles.}
\label{fig:G5_Comparison}
\end{figure}

\noindent The charmonium spectrum in ensemble B was measured omitting the quark-disconnected contributions to the correlation functions as a first test of the optimal profiles in a close-to-physical setup using $N_v = 325$ \cite{ThesisUrrea}. The lightest particle in this case is the $\eta_c$ with quantum numbers $J^{PC} = 0^{-+}$, accessible with $\Gamma = \gamma_5$. Figure \ref{fig:G5_Comparison} shows the effective masses of the ground state using $\Gamma = \gamma_5$ both with standard distillation and with the optimal profile from the GEVP. The suppression of excited-state contamination when using the optimal profile is clear; the mass plateau starts earlier which effectively more than doubles the plateau interval in this case. This not only yields a more reliable estimate of the plateau average but also means the signal at relatively small time separations is already dominated by the ground state. This is particularly important to extract a useful signal when quark-disconnected contributions are taken into account. 

\begin{figure}[H]
\centering
\includegraphics[width=0.7\textwidth]{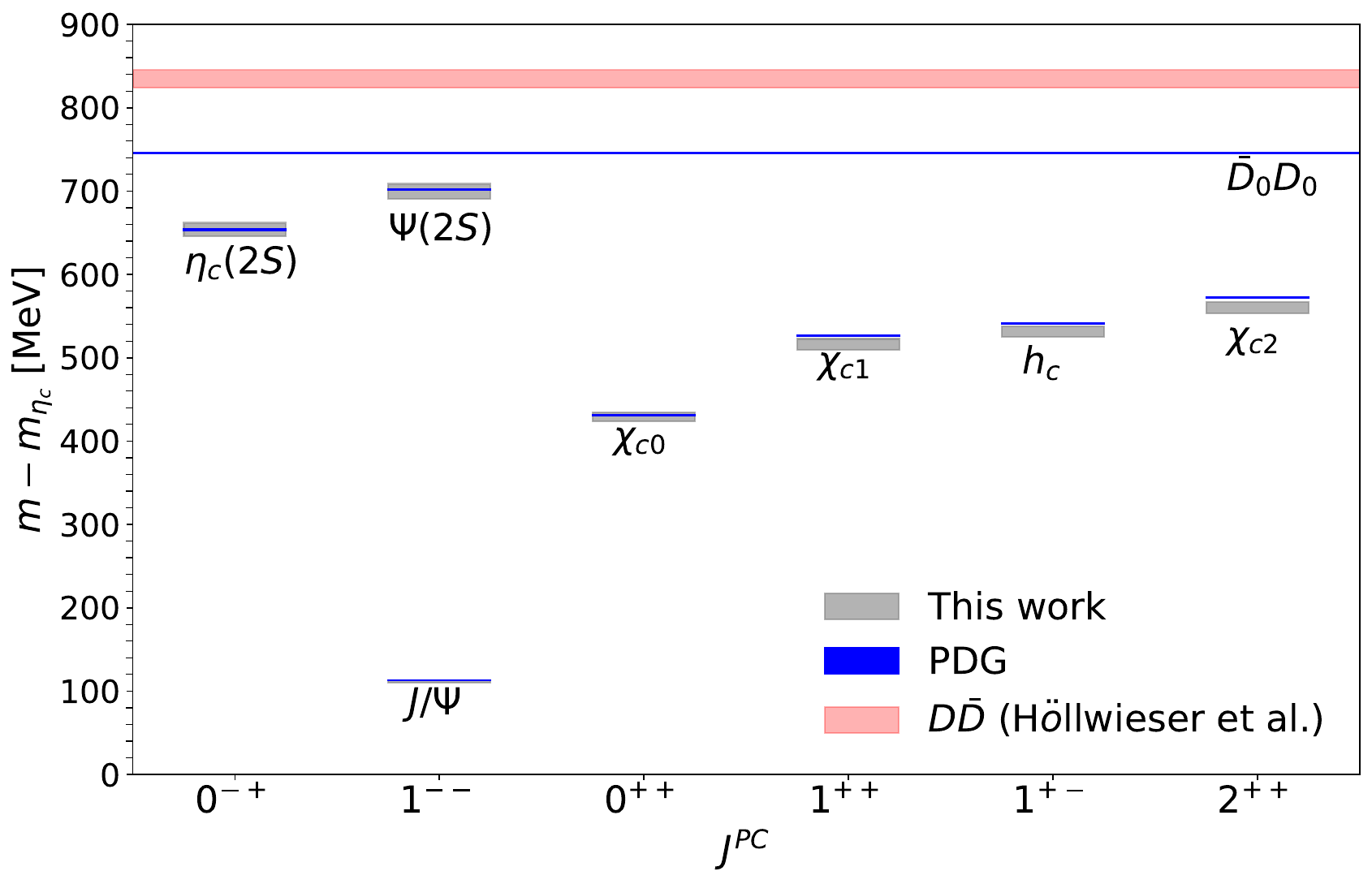}
\caption{Low-lying charmonium spectrum obtained using optimal meson distillation profiles from quark-connected correlation functions in \cite{ThesisUrrea} compared with the values from experiment..}
\label{fig:Spectrum_B}
\end{figure}

\noindent A similar improvement was obtained for $\Gamma$ operators corresponding to other lattice irreps and the resulting mass estimates for all states below the $D\bar{D}$ threshold are shown in Fig. \ref{fig:Spectrum_B}, where the continuum quantum numbers $J^{PC}$ are used as labels. The calculated mass of the $\eta_c$ was subtracted from all other masses to eliminate the effect of the charm quark mistuning. The gray rectangles correspond to these relative masses calculated in this work while the blue rectangles correspond to their value in nature \cite{Workman}. There is good agreement between the results in this work and the experimental counterparts even with the omission of quark-disconnected contributions, indicating these are most probably suppressed. In particular, the hyperfine splitting in this work ($111.8(1.4)$ MeV) lies very close to the experimental $113.0(5)$ MeV. It has a level of statistical uncertainty which is competitive with other state-of-the-art lattice calculations which do not use distillation, e.g $118.6(1.1)$ MeV \cite{Hatton} and $116.2(1.1)$ MeV \cite{DeTar}. The $D\overline{D}$ threshold for this ensemble is shown in red, using the mass of the $D$-meson measured in \cite{Roman-Ensembles}, while the experimental result of the $D_0 \overline{D}_0$ one is shown in blue. The difference between these two values indicates the effects of not having the light quark masses at their physical values.

\begin{figure}[H]
\centering
\includegraphics[width=0.65\textwidth]{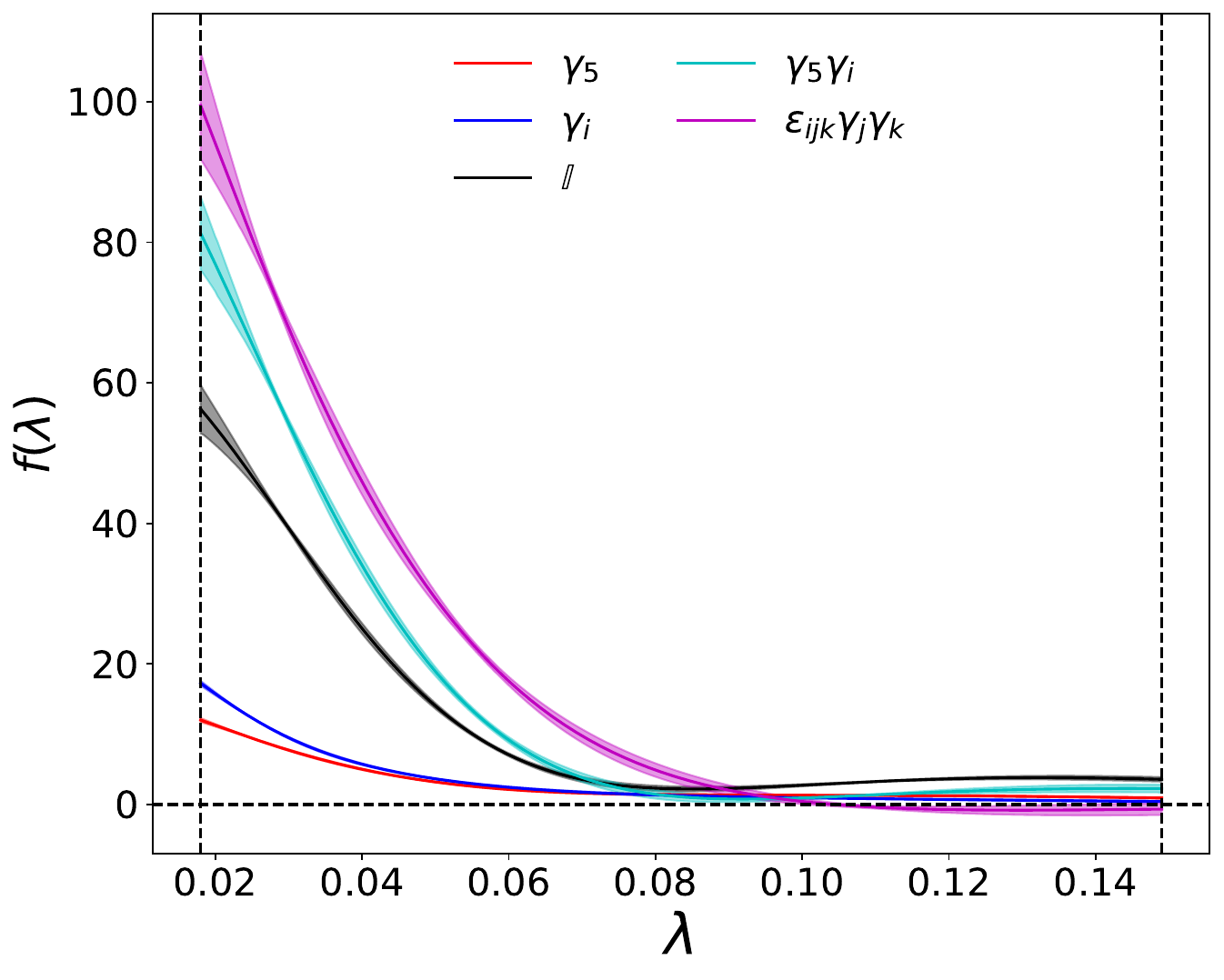}
\caption{Ground state optimal profiles for a selection of $\Gamma$ operators in charmonium.}
\label{fig:LocalProfiles}
\end{figure}

\begin{figure}[H]
\centering
\includegraphics[width=0.7\textwidth]{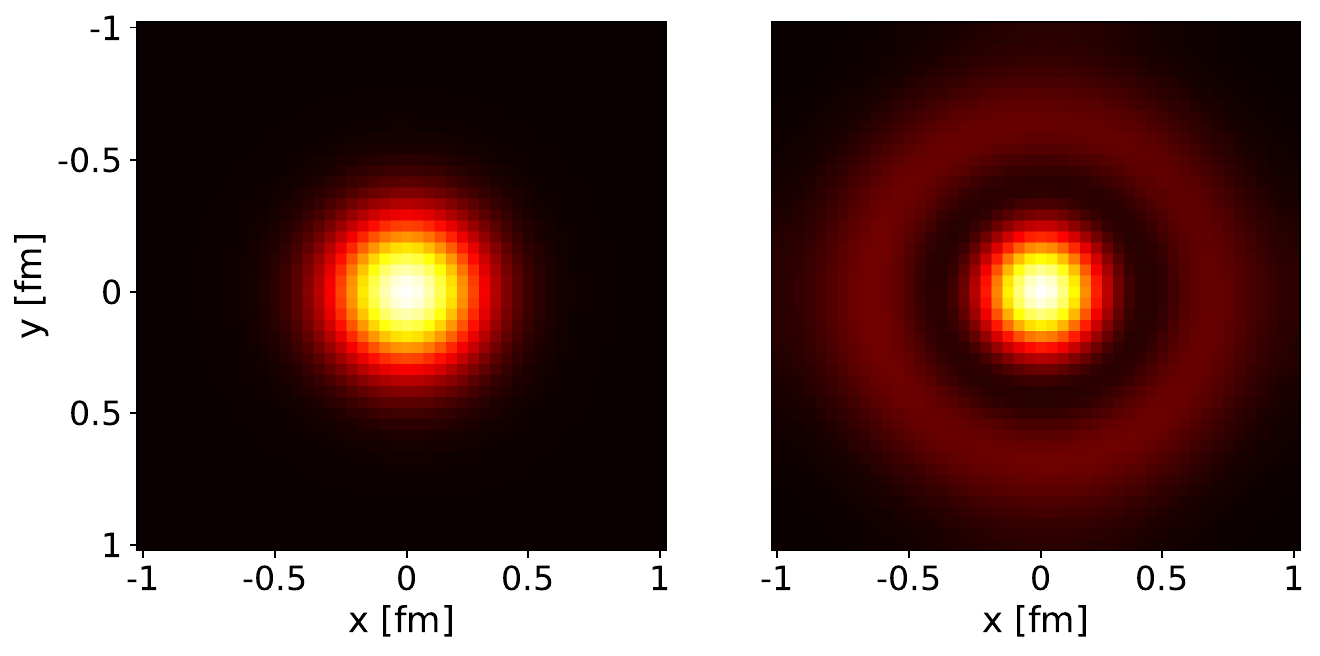}
\caption{Ground and first excited state spatial profiles of the $\Gamma = \gamma_5$ charmonium operator.}
\label{fig:SpatialProfiles}
\end{figure}

\noindent The optimal profiles for the ground state of $\Gamma$ operators based only on Dirac matrices which come from the same GEVP as the reported masses are shown in Fig. \ref{fig:LocalProfiles}. None of them resemble a constant, a feature already observed when the optimals profiles were first studied \cite{Urrea}. Higher Laplacian eigenvalues are significantly suppressed, so fewer eigenvectors could be used to extract these ground states. However, since excited states are also of interest it is worth to keep all the calculated eigenvectors. As presented in \citep{Urrea}, it is possible to define a spatial profile for spin-singlet operators involving $\gamma_5$ in their $\Gamma$ operator and the ground and first excited state spatial profiles for $\Gamma = \gamma_5$ are shown in Fig. \ref{fig:SpatialProfiles}. The expected S-wave behavior is observed in terms of radial structure and presence of nodes. The spatial size of the lattice provides a good resolution for the profiles and these seem to be well contained in the volume, indicating finite-volume effects are under control for these two states.

\begin{figure}[hpt]
\centering
\includegraphics[width=0.9\textwidth]{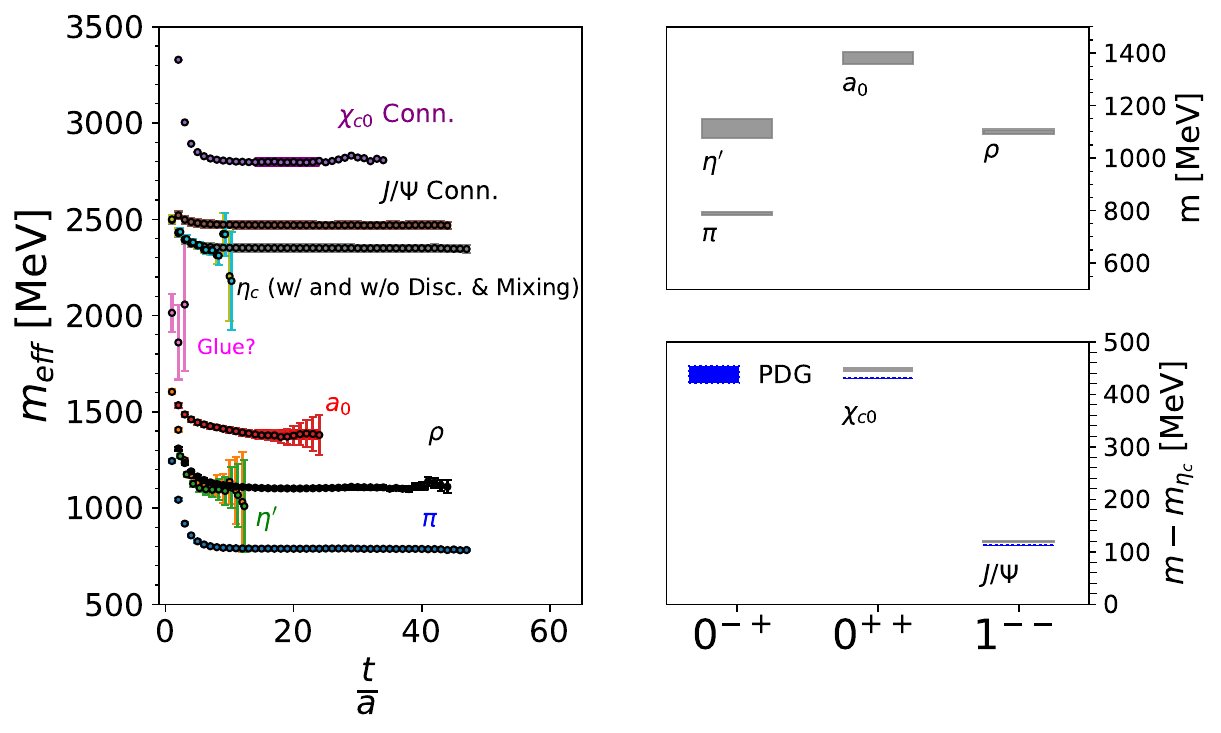}
\caption{Charmonium and light spectrum in ensemble A1-heavy.}
\label{fig:Spectrum_A1heavy}
\end{figure}

\begin{figure}[hpt]
\centering
\includegraphics[width=0.8\textwidth]{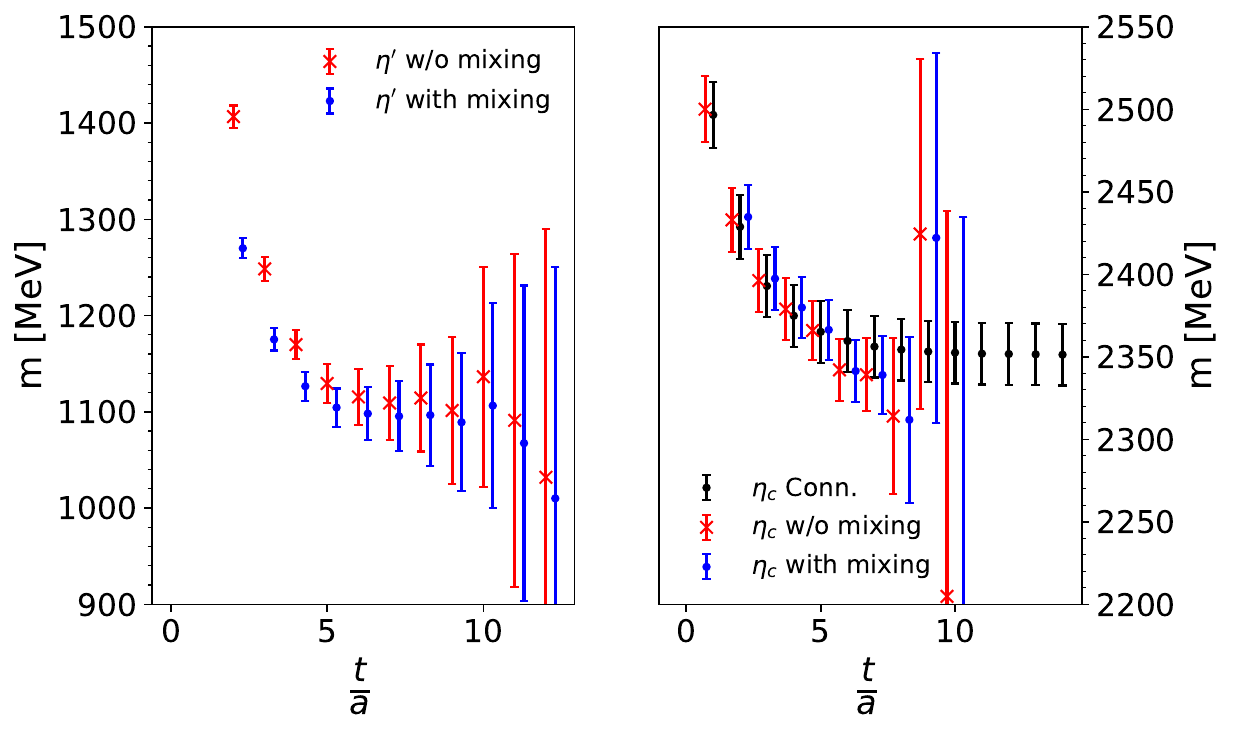}
\caption{Close-up of the effective masses of the $\eta^{\prime}$ and $\eta_{c}$. Red points neglect flavor-mixing between light and charm while blue points take it into account. Black points for the $\eta_c$ correspond to using only the connected contribution to the correlation functions.}
\label{fig:Results_A1heavy}
\end{figure}

\begin{figure}[H]
\centering
\includegraphics[width=.7\textwidth]{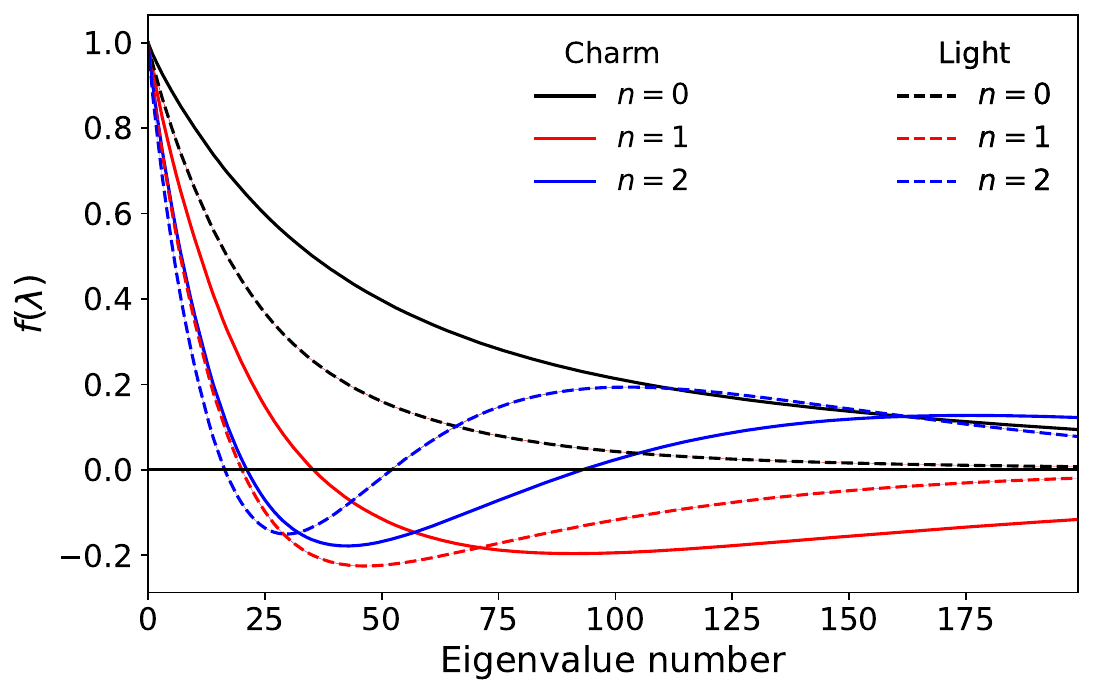}
\caption{Optimal meson distillation profiles for the $0^{-+}$ charmonium and light channels up to the second energy eigenstate.}
\label{fig:Profiles_Heavy}
\end{figure}

\noindent In ensemble A1-heavy both the charmonium and light meson spectrum were mapped using $N_v = 200$ for both quark masses. Fig. \ref{fig:Spectrum_A1heavy} shows the effective masses for different particles of interest together with the calculated mass plateau averages. To compare the charmonium values with experiment, the mass of the $\eta_c$ is subtracted from the masses of the $J/\Psi$ and $\chi_{c0}$ which eliminates the effects of the mistuning of the charm quark mass. The resulting masses show good agreement with experimental values \citep{Workman}. The channels involving only quark-connected contributions in the correlation functions display the clearest signal, e.g the pion, while the ones involving quark-disconnected contributions are affected by the signal-to-noise problem at very early times, e.g the $\eta^{\prime}$. The effective masses coming from the purely gluonic operators for the $0^{++}$ channel are also displayed and seem to approach a value slightly below $2$ GeV, close to the two-pion threshold. Nonetheless the error becomes too large to make a definitive statement and the inclusion of a two-pion operator is expected to help in this regard. Cases of particular interest are the $\eta^{\prime}$ and $\eta_c$, whose effective masses are more clearly displayed in Fig. \ref{fig:Results_A1heavy}. Their effective masses were calculated with and without taking into account the mixing between charmonium and light meson operators, therefore any differences between the points will be due to these dynamics. Since both these particles are in the same symmetry channel, they correspond to different energy eigenstates of the same $J^{PC}$. The ground state is the $\eta^{\prime}$ and the $\eta_{c}$ is higher up the ladder of excitations, some of which are two-particle states. The mass of the $\eta_c$ from only connected contributions is included for reference, assumed to be not far from the true iso-scalar state. Since the masses from the GEVP using disconnected contributions and mixing start very close to this reference point, it seems the charmonium operators have a large overlap with a state close to this reference. Nonetheless, the trend in the effective masses to go down before the error becomes too large indicates a non-zero overlap with lower states. Both with and without mixing, the $\eta^{\prime}$ masses remain consistent with each other and a reduction of excited-state contamination is seen in the case with mixing. Fig. \ref{fig:Profiles_Heavy} shows the profiles for the first three states of the $0^{-+}$ channel in charmonium and light mesons, where again the non-trivial structure in distillation space is observed. While there are similarities between the charm and light profiles in terms of number of nodes, the most prominent feature is the earlier suppression of eigenvalues in the light profiles compared to the charm ones. This indicates fewer eigenvectors are required for the study of the light spectrum, which can represent a significant reduction of computational costs since the light inversions are considerably more expensive than the charm ones.

\section{Conclusions}
\noindent This work showed that the use of optimal distillation profiles yields a significant improvement over standard distillation in the study of charmonium in a setup with a physical charm quark and three degenerate light quarks at the $SU(3)$ flavor symmetric point with two different pion masses. The charmonium spectrum with the lighter pion is in good agreement with experiment and the statistical uncertainty is compatible with other state-of-the-art lattice calculations. The spectrum of light mesons at two different pion masses was also mapped via this same method and the mixing between charm and light iso-scalar mesons was studied in the ensemble with a heavy pion. Small effects of charm-light flavor mixing were observed for the case of the $\eta^{\prime}$ and $\eta_{c}$ states. The optimal profiles for the light mesons are narrower than the charmonium ones, indicating that fewer eigenvectors are required to access the energy eigenstates of interest. This represents a significant reduction of computational costs, since the light inversions are more expensive than the charm ones. Some signal for a scalar glueball slightly below $2$ GeV was observed, close to the two-pion decay threshold in this setup, yet better gluonic and two-pion operators are required to perform a systematic study of these decay dynamics, which is a work in progress.\\

\textbf{Acknowledgement.} The authors gratefully acknowledge the Gauss Centre for Supercomputing e.V. (www.gauss-centre.eu) for funding this project by providing computing time on the GCS Supercomputer SuperMUC-NG at Leibniz Supercomputing Centre (www.lrz.de) under GCS/LS project ID pn29se as well as computing 
time and storage on the GCS Supercomputer JUWELS at Jülich Supercomputing Centre (JSC) under GCS/NIC project ID HWU35. The authors also gratefully acknowledge the scientific support and HPC 
resources provided by the Erlangen National High Performance Computing 
Center (NHR@FAU) of the Friedrich-Alexander-Universität 
Erlangen-Nürnberg (FAU) under the NHR project k103bf. M.P. was supported by the European Union’s Horizon 2020 research and innovation programme under grant agreement 824093 (STRONG-2020). R.H. is supported by the programme "Netzwerke 2021", an initiative of the Ministry of Culture and Science of the State of Northrhine Westphalia, in the NRW-FAIR network, funding code NW21-024-A. The work is supported by the German Research Foundation (DFG) research unit FOR5269 "Future methods for studying confined gluons in QCD". 

\bibliographystyle{JHEP}
\bibliography{refs} 

\end{document}